\documentclass[a4paper,10pt]{article}
\usepackage{amsfonts,latexsym,color}
\usepackage{amsmath}
\usepackage{amscd}
\usepackage{float,amsmath,amssymb,mathrsfs,bm,multirow,graphics}
\usepackage[dvips]{graphicx}

\def\b{\begin{eqnarray}}
\def\e{\end{eqnarray}}
\def\n{\noindent}

\newcommand{\C}{{\Bbb C}}

\newcommand{\Z}{{\Bbb Z}}

\begin{document}

\begin{center}

{\LARGE\textbf{Dressing Method for the Degasperis-Procesi Equation
\\}} \vspace {10mm} \vspace{1mm} \noindent

{\large \bf Adrian Constantin$^{a,\dag}$} and {\large \bf  Rossen I.
Ivanov$^{b,\ddag}$} 

\vskip1cm

\n

\hskip-.3cm
\begin{tabular}{c}
\hskip-1cm $\phantom{R^R} ^{a}${\it Department of Mathematics, King's College London, Strand, WC2L 2RS, London, UK,} \\
and {\it Faculty of Mathematics,
University of Vienna, Oskar-Morgenstern-Platz 1, 1090 Vienna, Austria} \\
\\
$\phantom{R^R}^{b}${\it School of Mathematical Sciences, Dublin Institute of Technology, Kevin Street}\\ {\it Dublin 8, Ireland} \\
\\
\\
\\{\it $^\dag$e-mail: adrian.constantin@univie.ac.at,}
\\{\it $^\ddag$e-mail: rossen.ivanov@dit.ie  }
\\
\hskip-.8cm
\end{tabular}
\vskip1cm
\end{center}



\vskip1cm

\begin{abstract}
\noindent The soliton solutions of the Degasperis-Procesi equations are constructed by the implementation of the dressing method. The form of the one and two soliton solutions coincides with the form obtained by Hirota's method. \end{abstract}

\section{Introduction}

In a previous publication \cite{CIL} we developed the inverse scattering transform (IST) for the Degasperis-Procesi (DP) equation by reducing it to an $\mathfrak{sl}(3)$ Zakharov-Shabat spectral problem with constant boundary
conditions. In \cite{CL} we constructed the fundamental analytic solutions of the spectral problem and we also provided a formulation of the IST as a Riemann-Hilbert problem. The construction of the solitons from the discrete spectrum has been outlined in principle by employing the well-known dressing method of Zakharov, Shabat and Mikhailov \cite{ZaSh1,ZS1,ZM,Sh75,Sh79,AV}, see also \cite{Ger,VSG1,JH,JC09,I3}. The explicit form of the solitons however has not been obtained. The DP solitons are known from the papers of Matsuno \cite{M1,M2}, where he used Hirota's method. Thus there remains the important question of the soliton derivation by the methods from spectral theory, such as the dressing method. This paper is aimed at addressing this question. Several significant papers on the spectral theory of the DP equation have appeared.
For example, in \cite{BMS1} it is shown that the Riemann-Hilbert formulation of the IST for the DP equation can used to study the long-time behaviuor 
of the solutions. In \cite{L2} the IST for the DP equation on a half-line is studied. 
In this sense the results of the present study are complementary to those in \cite{CIL,BMS1,M1,M2} and contribute to the completeness of the spectral theory of the DP equation, which is similar to the spectral theory of several other integrable equations, like the Kaup-Kuperschmidt equation \cite{Ka,TV08}, the Tzitzeica equations \cite{BCG1,BCG2} and the Sawada-Kotera equation \cite{SK}.  

By now the amount of literature on the DP equation is enormous --- the equation even has an entry on Wikipedia. Some of the basic facts for the DP equation can be found in \cite{CIL} and the references therein. The equation is usually written in the form
\begin{equation}\tag{DP}\label{DP}
  u_t - u_{txx} + 3\kappa u_x+4uu_x - 3u_xu_{xx} - uu_{xxx}=0,
\end{equation}
where $\kappa > 0$ is a constant. It was first discovered in \cite{D-P} in a search for asymptotically integrable PDEs, which have nonlinear terms similar to the famous integrable Camassa-Holm (CH) equation \cite{C-H, FF, F, CFHT}. As a matter of fact, in this family of equations there are no other integrable representatives (see \cite{CL, I}). The Lax pair and other important structures for \eqref{DP} are reported in \cite{D-H-H, HW}. Like the CH equation, \eqref{DP} is also a water-wave equation \cite{J, CL, I2,HI}.
A particular feature of the DP equation is that a stronger nonlinearity could allow for the occurence of wave-breaking --- a fundamental phenomenon in the theory of water waves that is not captured by other models like the KdV equation (see the discussion in \cite{CE}). 

The DP equation also admits peakon' solutions \cite{D-H-H, HW, M2,D-H-H2, L-S,CHT}. They can be obtained in the limit $\kappa \to 0$ from the solitary waves \cite{M2} and are given explicitly by
$$u_c(x,t)=c\,e^{-|x-ct|},\qquad x,\,t \in {\mathbb R},$$
where the constant $c>0$ represents the wave speed. Such solutions with a peak at the wave crest of course have to be understood as weak solutions \cite{L}. Their shape resembles that of the celebrated `Stokes wave of greatest height' (see the discussion in \cite{C, CE2}). Note that the 
peakon solutions are orbitally stable --- their shape is stable under small perturbations (see \cite{CS, LL}) --- and therefore these wave 
patterns are detectable.

In this paper, we will apply the dressing method approach to smooth
localized solutions to (\ref{DP}) in order to obtain the smooth regular solitons. More precisely, we consider solutions $u(x,t)$ of class $C^1$ in
$t$ and of Schwartz class regularity with respect to the $x$-variable (i.e. the solution is smooth and decays to zero faster than any polynomial as $|x| \to \infty$). Moreover, we assume that the solution satisfies the following inequality initially (at time $t=0$):
\begin{equation}\label{q}
q = u- u_{xx} + \kappa > 0.
\end{equation}
Note that well-posedness and global existence of solutions
for (\ref{DP}) holds within the class of Schwartz functions if the initial data satisfy (\ref{q}), and in this case the validity of (\ref{q}) is ensured at any later time $t > 0$ (see \cite{ELY, H, LY}).

The paper is organised as follows. In Section 2 we present some general facts about the Lax pair formulation of (\ref{DP}) and the symmetry properties of the isospectral problem. In Section 3 we construct the one- and two-soliton solutions by applying the dressing method.

\section{Spectral problem}

\subsection{Lax pair}

The Lax pair for the equation (\ref{DP}) is of third order in the $x$-derivatives \cite{D-H-H}, in contrast to the Lax pair for the CH equation, which has a second order spectral problem \cite{C-H, CGI, BMS2, BMS3, CM}.
%

The Lax representation can be written in the form
of a Zakharov-Shabat (ZS)-type matrix spectral problem \cite{CIL}:
\begin{equation}\label{matrixlax}
\begin{cases}
\phi_{x} = \tilde{L}\phi, \\
\phi_t = \tilde{M} \phi,
\end{cases}
\end{equation}
where
$$\tilde{L} = \begin{pmatrix}  -1 & \zeta  & 0 \\
 0 & 0 & \zeta  \\
 \zeta  q & 0 & 1 \end{pmatrix},
 $$
 
 $$
 \tilde{M} = \begin{pmatrix}
  u+u_x + \frac{1}{3 \zeta^3} & -\zeta u-\frac{1}{\zeta^2} & \frac{1}{\zeta } \\
  \frac{q+u_x+u_{xx}}{\zeta } & -\frac{2}{3 \zeta ^3} & \frac{1}{\zeta ^2}-\zeta  u \\
 \frac{-q u \zeta^3 + q_x - u_x + u_{xxx}}{\zeta ^2} & \frac{q + u_x + u_{xx}}{\zeta }
 & -u - u_x + \frac{1}{3 \zeta^3}
   \end{pmatrix}
 $$
$\zeta \in \C$ is the spectral parameter and $\phi(x,t)$ is a $SL(3)$ - matrix-valued function, whose
columns, considered as vectors, represent the three linearly
independent solutions of the matrix equation. This allows us to take advantage of the existing inbuilt symmetries of the considered equation. More about the ZS system, the AKNS system \cite{AKNS} and their generalisations can be found in \cite{ZS1,Sh79,Sh75,ZMNP,FaTa,GVY,Ger,VSG86,VSG1}.

Let $G(x,t)$ be a $SL(3)$ matrix and let us perform the change of variables $\phi = G \psi$. It transforms (\ref{matrixlax}) into
$$\begin{cases}
\psi_{x} = L\psi, \\
\psi_t = M \psi,
\end{cases}$$
where
$$L = G^{-1} \tilde{L} G - G^{-1} G_x, \qquad M = G^{-1} \tilde{M} G - G^{-1} G_t\,,$$
and
$$G = \begin{pmatrix}
 q^{-1/3} & 0 & 0 \\
 0 & 1 & 0 \\
 0 & 0 & q^{1/3} \end{pmatrix}\,.$$
 we find \footnote{Here we are using notations slightly different from those in \cite{CIL}. The spectral problem is gauge-equivallent to the one in \cite{CIL}.} $L = \zeta q^{1/3} J - \tilde{Q}$ where $\tilde{Q}=\left(1 - \frac{q_x}{3q}\right)H_0$,
 \begin{equation}\label{Jqdef}
 J = \begin{pmatrix} 0 & 1 & 0 \\
 0 & 0 & 1 \\
 1 & 0 & 0\end{pmatrix} \quad \hbox{and}\quad
 H_0 = \begin{pmatrix}  1 &  0 & 0 \\
 0 & 0 & 0 \\
 0 & 0 & -1\end{pmatrix} .
 \end{equation}

 Let us change the variables according to

\b \label{y} y= x+\int_{-\infty} ^{x} \Big[
\Big(\frac{q(x')}{\kappa}\Big)^{1/3}-1 \Big]\text{d}x', \qquad
\frac{\text{ d} y }{\text{ d} x}
=\Big(\frac{q(x)}{\kappa}\Big)^{1/3}.\e

\n  The $t$-variable can be viewed as an additional parameter
rather than a second independent variable. This will be clear in
the following section and is due to the fact that the
$t$-dependence of the scattering data can be explicitly computed
in relatively simple form. For the sake of simplicity, in what
follows, we usually omit the $t$-dependence of the variables,
unless this dependence is necessary for the computations. The
spectral problem
\begin{equation}\label{newxpart}
  \psi_x + (\tilde{Q} - \zeta q^{1/3} J ) \psi = 0
\end{equation}

\n can be written in the form

\begin{equation}\label{SP}
\psi_y + (h H_0- \lambda J ) \psi = 0,
\end{equation} where $h$ is a scalar function, \b \label{h} 
h(x)=\Big(\frac{q(x)}{\kappa}\Big)^{-1/3}+\frac{\text{ d} }{\text{
d} x}\Big(\frac{q(x)}{\kappa}\Big)^{-1/3}, \qquad \lambda=\zeta \kappa^{1/3} \e

\n and $\lim_{x\rightarrow \pm \infty} h(x)=1$.

Suppose that $x=X(y)$. It is possible to recover $q(x)$ from
$h(X(y))$, see \cite{CIL}. We notice that asymptotically $y\rightarrow x$
when $x\rightarrow - \infty$. Since
$$\int_{-\infty} ^{\infty}\Big[ \Big(\frac{q(x)}{\kappa}\Big)^{1/3}-1 \Big]\text{d}x$$
is an integral of motion \cite{D-H-H}, for $x\rightarrow \infty$ we have that $x$
and $y$ differ only by a constant. Thus  $\lim_{y\rightarrow \pm \infty} h(y)=1$. 

\subsection{Automorphisms and reductions}

The specific form of the spectral problem in (\ref{SP}) is due to the
symmetry of the problem under the action of three distinct
automorphisms. In other words, the fact that the solution is determined by
a single real (scalar) function $h$, rather than 6 complex functions
(which is the case for an arbitrary $sl(3)$ potential) is a
consequence of its invariance under one $\Z_3$ automorphism and
two $\Z_2$ automorphisms. The automorphisms lead to the reduction
of the independent components of the matrix entries of $L$, $M$ and their action extends to the spectrum and the eigenfunctions. They form a group,
known as \emph{a reduction group}
\cite{AV}, see also \cite{GKV,GVY,GGK05a,GG1}.

\subsubsection{$\Z_3$ automorphism} The spectral problem (\ref{SP})
has a manifest $\Z_3$ symmetry (invariance):

\b C^{-1} L(\omega^2\lambda)C = L(\lambda), \label{Z3}\e

\n where $\omega=e^{2\pi i/3}$ and
$$C =  \begin{pmatrix}
1 & 0 & 0 \\
0 & \omega & 0 \\
0 & 0 & \omega^2 \end{pmatrix}. $$
Indeed, one can check that
$$C^{-1}JC = \omega  J, \qquad C^{-1}H_0C = H_0\,,$$
\n from where (\ref{Z3}) follows immediately. Furthermore, one can verify that this invariance holds for all elements of the associated graded Lie-algebra \cite{Ger}, including
\begin{equation}\label{Msymmetry}
  C^{-1} M( \omega^2 \lambda)C = M( \lambda),
\end{equation}
The $\Z_3$ invariance  (\ref{Z3}) 
applies to the group-valued solutions:

\b C^{-1}\psi ( y,t,\omega^2\lambda) C=\psi(y,t, \lambda) \label{A1} \e 
Similar type relations hold for all automorphisms of the spectral problem.

\subsubsection{$\mathbb{Z}_2$ automorphisms}

The spectral problem possess two additional $\mathbb{Z}_2$
automorphisms, one of which reflects the reality of $h(y)$. The
first one leads to  \b  \overline{\psi(y,t,\overline{\lambda})} =
\psi(y,t,\lambda), \label{Z2}\e

\noindent where the overline means complex conjugation of the quantity underneath.

\noindent
The second $\Z_2$ automorphism leads to the following symmetry of the group-valued eigenfunction:

\b \Gamma^{-1} \psi^{\dag}(y,t, -\bar{\lambda})\Gamma = \psi^{-1}(y,t,\lambda), \label{Z2-2}\e

\noindent where $\Gamma^{-1} H_0^{\dag}\Gamma =-H_0$ and $\Gamma^{-1} J^{\dag}\Gamma =J$ with

$$\Gamma =  \begin{pmatrix}
0 & 0 & 1 \\
0 & 1 & 0 \\
1 & 0 & 0\end{pmatrix}.
$$
Here the dagger stands for a matrix Hermitian conjugation.

One can also check that all $\Z_2$ and $\Z_3$ automorphisms are
compatible, in the sense that the order of their application to
quantities like $L$, $M$, $\psi$  does not matter.

\subsection{Diagonalisation}

We call the spectral problem and the associated quantities 'naked' if they are evaluated for the 'trivial' solution $u(x,t) \equiv 0$, or $h\equiv 1$:
$$L_0 = \lambda  J - H_0.$$
$L_0$ and $M_0$ are the `naked' matrices, and we note that, due to $\lim_{y \to \infty}u(x(y,t),t)=0$, the asymptotic values are 
$$L_\infty (\lambda) \equiv \lim_{y \to \pm \infty} L(y,\lambda)=L_0 (\lambda), \qquad M_\infty (\lambda)\equiv \lim_{y \to \pm \infty} M(y,\lambda)=M_0(\lambda),$$
\n i.e.
$$L_\infty = L_0=\lambda  J - H_0.$$
We also find that \b
M_0 (\lambda)=\frac{ \kappa}{3\lambda ^3}
\begin{pmatrix}
1 &-3\lambda    &3\lambda^2  \\
 3\lambda^2   & -2 &  3\lambda  \\
 0 & 3\lambda^2  & 1
   \end{pmatrix} \label{M_infty}\e  
   with
$$[L_0 , M_0 ] = 0.$$
Since $L_0$ and $M_0$ commute, they can be
simultaneously diagonalized. Let $U(\lambda)$ be a $SL(3)$ matrix
such that
$$L_0 (\lambda)  = U (\lambda)  \Lambda (\lambda) U^{-1}(\lambda) ,
\qquad M_0(\lambda)  = U(\lambda)  A(\lambda) U^{-1}(\lambda)
,$$
with
$$
\Lambda(\lambda)  = \text{diag}(\Lambda_1(\lambda) ,
\Lambda_2(\lambda) , \Lambda_3(\lambda) ),\qquad
A(\lambda) =\text{diag}(A_1(\lambda) , A_2(\lambda) , A_3(\lambda) ),$$
where $\Lambda_1, \Lambda_2, \Lambda_3$ and $A_1, A_2, A_3$ are the
eigenvalues of $L_0$ and $M_0$ respectively.

The eigenvalues of $L_{0}(\lambda)$  are the
solutions $\Lambda(\lambda)$ of the characteristic equation
\b \Lambda^3-\Lambda -\lambda^3 = 0. \label{cubiceq}\e

\n Introducing a new spectral parameter $k$ such that \b
\lambda(k)=3^{-1/2}k\Big(1+\frac{1}{k^6}\Big)^{1/3}=3^{-1/2}k\left(1+\frac{1}{3k^6}+\ldots\right),
\label{lambda(k)}\e we obtain \b
\lambda^3=3^{-3/2}\Big(k^3+\frac{1}{k^3} \Big)\e and the following
expression for the eigenvalues of $L_{0}$: \b \label{Lambdas}
\Lambda_j (k)=3^{-1/2}\Big(\omega^j k+\frac{1}{\omega^j k}\Big).
\e

\n Furthermore, $\lambda(k)$ has the property $\lambda(\omega
k)=\omega \lambda(k)$ and also \b \lambda(k)\rightarrow 3^{-1/2}k
\qquad \text{when} \qquad |k| \rightarrow \infty. \e

The characteristic polynomial of the matrix $3\kappa^{-1}\lambda
^3 M_{0}$,  cf. (\ref{M_infty}), is \b P(w)=w^3-3w-27
\lambda^{6} +2, \label{char poly M} \e

\noindent with roots
$$w_{j}(k)=\omega^{j}k^2+\omega^{-j}k^{-2},$$
where $\lambda(k)$ is given in (\ref{lambda(k)}). Thus the
eigenvalues of $M_{0}$ are (in accordance with  \cite{CIL})

\b A_j(k)&=&\sqrt{3} \kappa \frac{(\omega^j k)^{2}+(\omega^j
k)^{-2}}{k^3+k^{-3}}.  \label{mu_n_k} \e

\section{Dressing method}

The dressing method allows the explicit construction of a
solution with discrete eigenvalues $\psi(y,t,k)$, starting from a
trivial, or 'naked' solution of the spectral problem $\psi_0(y,t,k)$: \b \psi(y,t,\lambda(k))=g(y,t,\lambda(k))\psi_0(y,t ,\lambda(k))
\label{dressSolution}.\e 
The dressing factor $g$ is analytic in the entire complex $\lambda$ plane, with the exception of the points of the discrete spectrum. 

We make
the following assumptions in our construction of a dressing
factor.  First, we allow only simple poles of $g$ and $g^{-1}$.
The eigenfunctions, and the dressing factors are invariant under the corresponding automorphisms, thus $g(\lambda)$ is a group element and therefore:

\b C^{-1} g(y,t,\omega^2 \lambda)C&=&g(y,t, \lambda), \label{S1} \\
 \bar{g}\left(y,t, \bar{\lambda}
\right)&=&g(y, \lambda), \label{S2} \\
\Gamma^{-1} g^{\dag}(y,t,- \bar{\lambda})\Gamma&=&g^{-1}(y,t, \lambda),
\label{S3} \e

From these symmetries it follows that if $g$ or $g^{-1}$ have a
pole at, say, $\lambda_0$, then they have also poles at
$-\lambda_0$, $\pm \omega \lambda_0$, $\pm \omega^2 \lambda_0$,
$\pm \bar{\lambda}_{0}$, $\pm \omega \bar{\lambda}_{0}$, $\pm
\omega^2 \bar{\lambda}_{0}$. It is also possible for $g$ to have only 3 poles at  $\lambda_0$, $\omega \lambda_0$ and $\omega^2 \lambda_0$ and then, due to the symmetries, $g^{-1}$ will have poles at  $-\bar{\lambda}_0$, $-\omega \bar{\lambda}_0$ and $-\omega^2 \bar{\lambda}_0$. 
Note that since $g$ is in the Lie group, $g^{-1}$ should exist everywhere, i.e. $\det(g)\ne 0$.

\subsection{One-soliton solution}

The one-soliton solution corresponds to one discrete eigenvalue, $\lambda_1$. Let us assume that $\lambda_1$ can be chosen real, so that the following choice of $g$ is possible

\b
g(y,t,\lambda)=\!\!1\!\!\text{I}+\frac{1}{3}\left(\frac{\mathcal{A}_1(y,t)}{\lambda-\lambda_1}+\frac{C^{-1}\mathcal{A}_1(y,t) C}{\omega^2\lambda-\lambda_1}+\frac{C^{-2}\mathcal{A}_1(y,t)C^2}{\omega \lambda - \lambda_1} \right),\label{gfactor} \e

\n for some residues determined by real $\mathcal{A}_1(y,t)$. With this Anzatz for the dressing factor, the conditions \eqref{S1} and \eqref{S2} are automatically satisfied. The condition \eqref{S3} necessitates

\b g(y,t,\lambda)\Gamma^{-1} g^{\dag}(y,t,- \bar{\lambda})\Gamma=\!\!1\!\!\text{I}
\label{S3a} \e
 
This leads to the following equation for $A_1$ :
\b
\left(\!\!1\!\!\text{I}-\frac{1}{3}\left(\frac{\mathcal{A}_1(y,t)}{2\lambda_1}+\frac{C^{-1}\mathcal{A}_1(y,t) C}{(\omega^2+1)\lambda_1}+\frac{C^{-2}\mathcal{A}_1(y,t)C^2}{(\omega +1) \lambda_1} \right)\right)\Gamma^{-1}\mathcal{A}_1^T \Gamma =0,\label{A1eq} \e which is the condition of the vanishing of the residue at $-\lambda_1$ of \eqref{S3a}. The conditions at the other residues are equivalent due to the action of the reduction group.

We seek a solution of the matrix equation \eqref{A1eq}
(see also \cite{ZM,Ger,BCG1,BCG2}), in the form \b \mathcal{A}_1=|n\rangle \langle m| \label{A1nm}\e

\n where $|n \rangle $ is a 3-component vector-column, and  $\langle m|$
is a 3-component vector-row. It is a straightforward exercise to get the following relation for the components of the two vectors (indexed as usual with indices from 1 to 3):

\begin{align} 
n_1 & =\frac{2\lambda_1 m_3}{2m_1m_3-m_2^2} , \label{mn} \\
n_2 & =\frac{2\lambda_1}{m_2}  \label{mn2} , \\
n_3 & = \frac{2 \lambda_1 m_1}{m_2^2} \label{mn3}.  
\end{align}

The equation for the dressing factor $g$, that follows from the fact that $\psi$ from \eqref{dressSolution} is a solution of \eqref{SP} (and $\psi_0$ is a 'naked' solution coresponding to $h=1$), is  \b g_y+h H_0 g-g H_0-\lambda [J,g]=0, \label{Eq4g factor}\e satisfied
identically for any $\lambda$. This equation leads to a differential equation for $\mathcal{A}_1$, from (\ref{gfactor}), and therefore to differential equations for the vectors ere $|n \rangle $ and  $\langle m|$, which are

\begin{equation}
\begin{split} (\langle m|)_y -\langle m|(H_0- \lambda_1 J)&=0, \\
(|n\rangle)_y + (h H_0- \lambda_1 J)|n\rangle &= 0.\end{split} \end{equation}

This is a remarkable result, showing that when $|m \rangle $ satisfies a 'naked' equation for a matrix operator $-L_0^{T}$ at $\lambda=\lambda_1$:
\b (|m \rangle)_y =(H_0- \lambda_1 J)^{T} |m \rangle,  \e
\n or 
\begin{equation}
\begin{split}
m_{1,y}=& m_1 - \lambda_1 m_3 , \\
m_{2,y}=&- \lambda_1 m_1 , \\
m_{3,y}= & - \lambda_1 m_2 - m_3.
\end{split}
 \label{meq}
\end{equation}

\n then  $|n\rangle$ is an eigenfunction of the spectral problem for the 1-soliton case with a discrete eigenvalue $\lambda=\lambda_1$. Thus, we obtain explicitly the components $m_k$ and then from the algebraic relations \eqref{mn} we determine the unknown $n_k$. Indeed, if $V(k)$ is the matrix, diagonalizing $-L_0^T$ and $\lambda_1=\lambda(k_1)$, clearly we can write 
$$|m \rangle = V(k_1) \mathrm{diag}(e^{-\Lambda_1(k_1) y - A_1(k_1) t},e^{-\Lambda_2(k_1) y - A_2(k_1) t},e^{-\Lambda_3(k_1) y - A_3(k_1) t}) |\mu \rangle,$$

\n where $|\mu \rangle$ is another arbitrary 3-component vector with components $\mu_j$. Computing $V(k_1)$ explicitly, we have
\begin{equation} \begin{split}
m_1&=\frac{1}{\lambda_1 }\sum_{j=1}^{3}\mu_j\Lambda_j(k_1)e^{-\Lambda_j(k_1)y-A_j(k_1)t} , \\
m_2&= \sum_{j=1}^{3}\mu_j e^{-\Lambda_j(k_1)y-A_j(k_1)t} , \\
m_3&=\lambda_1 \sum_{j=1}^{3}\frac{\mu_j}{\Lambda_j(k_1)-1}e^{-\Lambda_j(k_1)y-A_j(k_1)t} \label{m} .\end{split} \end{equation}
Thus, due to \eqref{mn}-\eqref{mn3}, we know $\mathcal{A}_1$ and the dressing factor $g$. 
The next step is to recover the solution $u(x,t)$. It is sufficient to obtain explicitly the change of variables $x=X(y,t)$. Then, in parametric form, 
we have that $u(X(y,t),t)=X_t(y,t)$, cf. \cite{M1, M2}. Recall that when $\lambda=0$ the equation in $x$ for $\psi$ is $$ \psi_x +\left(1 - \frac{q_x}{3q}\right)H_0 \psi =0 ,$$  with a solution 
\begin{equation}
\begin{split} \psi(x,t,\lambda=0)&=\exp(-H_0 ( x- \frac{1}{3}\ln (q/\kappa))\\
&=\mathrm{diag}\left(\left(\frac{q}{\kappa}\right)^{1/3}e^{-x}, 1,\left(\frac{q}{\kappa}\right)^{-1/3} e^x\right). \end{split} \label{psi01}\end{equation}

On the other hand, \b \psi(x,t,\lambda=0)=g(y,t,\lambda=0)\psi_0(y,t,\lambda=0)=g(y,t,\lambda=0)e^{-H_0 y} \label{psi02}\e The computation of $g(y,t,\lambda=0)$ produces

\b g(y,t,\lambda=0)=\mathrm{diag} \left(1-\frac{n_1m_1}{\lambda_1},1-\frac{n_2m_2}{\lambda_1}, 1-\frac{n_3m_3}{\lambda_1} \right) .  \e

Note that from \eqref{mn2} $n_2 m_2 = 2 \lambda _1$, so that

\b g(y,t,\lambda=0)=\mathrm{diag} \left(1-\frac{n_1m_1}{\lambda_1},-1, 1-\frac{n_3m_3}{\lambda_1} \right)   \e

The two solutions \eqref{psi01} and \eqref{psi02} must coincide up to an overall numerical constant, which, looking at the second diagonal entry, must be $-1$. Therefore, a comparison of the $33$-elements gives \b \left(\frac{q}{\kappa}\right)^{-1/3}  e^{x-y}=-\left( 1-\frac{n_3m_3}{\lambda_1}\right) \label{link}\e

\n Since $\left(\frac{q}{\kappa}\right)^{-1/3}=\frac{\partial X}{\partial y} $ from \eqref{y}, we arrive at an equation for $X(y,t)$:

\b  e^{X-y}\frac{\partial X}{\partial y}=-\left( 1-\frac{2m_1(y,t)m_3(y,t)}{m_2^2}\right) \e
The formal integration of the above equation is straightforward (separation of variables), however we can  provide an explicit solution in the form \b X(y,t)=y+ \ln \left(1+ \frac{2m_3}{\lambda_1 m_2} \right). \label{Xy}\e

\n This can be easily verified using the equations \eqref{meq} for $m_j$. Thus, \eqref{Xy} (together with $u(X,t)=X_t$) is a solution in parametric form (with a parameter $y$) of the one-soliton DP equation in terms of the scattering data, $\lambda_1$, and seemingly 3 other constants $\mu_j$:

\b X(y,t)=y+ \ln \left(\frac{\sum_{j=1}^{3}\frac{\Lambda_j+1}{\Lambda_j-1}\mu_j e^{-\Lambda_j y-A_j t}}{\sum_{j=1}^{3}\mu_j e^{-\Lambda_j y-A_j t}}\right)_{k=k_1}. \label{Xy1}\e

\noindent This expression formally produces smooth solitons for the DP equation
$u(X(y,t),t)$ in the variables $(y,t)$ for a wide range of the involved parameters. However, the requirement that $X(y,t)$ must be strictly monotone in $y$
gives additional restrictions on the parameters. These restrictions will be investigated in more details below.

 Apart from the eigenvalue $\lambda_1$, we expect only one additional constant (from the general theory of action-angle variables). Let us take a closer look at the solution \eqref{Xy1}. $\Lambda_j(k_1)$ are roots of the cubic equation \eqref{cubiceq}  with $\lambda=\lambda_1$, and therefore at least one root, say $\Lambda_3(k_1)$, is real.    
Since $\Lambda_3 (k_1)=3^{-1/2}\Big(k_1+\frac{1}{k_1}\Big).$ there are 2 options:

(i) $k_1$ real, $k_1=\pm e^{l_1}$ for some real $l_1\ne 0$. The case $l_1=0$ will be analysed separately.  Then  \b \Lambda_{1,2}(l_1) = -\frac{1}{2}\Lambda_3(l_1) \pm i \sinh(l_1) \e are complex conjugate. The reality of the expression under $\ln $ in \eqref{Xy1} can be insured by $\mu_1=\bar{\mu}_2$, $\mu_3$ - real. In this case, however, both the nominator and the denominator under  $\ln $ in \eqref{Xy1} contain oscillatory $\sin$ and/or $\cos$ terms (irrespective of the possible choices for $\mu_k$ ) and the function \eqref{Xy1} is not monotonic (it also develops singularities), i.e. it is not a valid change of variables. Thus we should rule out this option.

(ii)  $k_1$ complex. Since $\Lambda_3 (k_1)=3^{-1/2}\Big(k_1+\frac{1}{k_1}\Big)$ is real, $\frac{1}{k_1}=\bar{k}_1$ so that $|k_1|=1$ and $k_1=e^{il_1}$ for some real $l_1$. Let us assume that $l_1\neq 0$. Then $\Lambda_{1,2}$ are also real:

\b \Lambda_{1}(l_1) &=& -\frac{1}{\sqrt{3}}\cos(l_1)- \sin(l_1) ,\label{L1}\\
\Lambda_{2}(l_1) &=& -\frac{1}{\sqrt{3}}\cos(l_1)+ \sin(l_1) ,\label{L2}\\
\Lambda_{3}(l_1) &=& \frac{2}{\sqrt{3}}\cos(l_1) .\label{L3}\e

The positivity of the denominator in  \eqref{Xy1} can be ensured by choosing all $\mu_j$ real and positive. There is another issue with the positivity of the nominator: even though  $\mu_j$ can be chosen real and positive, the multipliers $\frac{\Lambda_j(k_1)+1}{\Lambda_j(k_1)-1}$  can not be all simultaneously positive, irrespective of $k_1$ or $l_1$: one can show that their product is always negative, for any choice of $l_1$. Therefore, at least one of $\mu_j$ should be chosen equal to zero, e.g. $\mu_3=0$, in order to eliminate the corresponding negative term $\frac{\Lambda_3(k_1)+1}{\Lambda_3(k_1)-1}$. Then of course  \eqref{Xy1} depends only on the ratio $\mu_1/\mu_2$, which is the second constant (related to the 'angle' variable),  in addition to the `action' variable $l_1$. This way all terms under $\ln$ are positive and well defined. The $A_j$ eigenvalues are

\b A_{1}(l_1) &=& \frac{\sqrt{3}\kappa}{\cos(3l_1)}\left(-\frac{1}{2}\cos(2l_1)+\frac{\sqrt{3}}{2} \sin(2l_1)\right) ,\\
A_{2}(l_1) &=&\frac{\sqrt{3}\kappa}{\cos(3l_1)}\left(-\frac{1}{2}\cos(2l_1)-\frac{\sqrt{3}}{2} \sin(2l_1)\right) ,\\
A_{3}(l_1) &=& \sqrt{3}\kappa\frac{\cos(2l_1)}{\cos(3l_1)}. \e

With these assumptions and introducing the constants $\gamma_1$ and $\sigma$ as shown below, \eqref{Xy1} can be written in the form

\b
x=X(y,t)&=&y+\ln \left( \frac{\gamma_1+1+(\gamma_1-1)e^{\xi_1}}{\gamma_1-1+(\gamma_1+1)e^{\xi_1}}\right)+\ln \sigma, \label{Xy2}\e
where the following relations hold 
\b \xi_1&=&2\sin(l_1) y-\frac{3\kappa \sin(2l_1)}{\cos(3l_1)}t
-\ln\left(\frac{\mu_2}{\mu_1}\sqrt{\frac{(\Lambda_1-1)(\Lambda_2+1)}{(\Lambda_1+1)(\Lambda_2-1)}}\right)_{l=l_1} \label{rad1} \e 

\b  \frac{\gamma_1+1}{\gamma_1-1}
&=&\left(\sqrt{\frac{(\Lambda_1-1)(\Lambda_2+1)}{(\Lambda_1+1)(\Lambda_2-1)}}\right)_{l=l_1} \label{gamma1},\\
\sigma&=&\left(\sqrt{\frac{(\Lambda_1+1)(\Lambda_2+1)}{(\Lambda_1-1)(\Lambda_2-1)}}\right)_{l=l_1}
\e

\n  Recall that the choice of $l_1$ is such that the two factors $\frac{\Lambda_{1,2}+1}{\Lambda_{1,2}-1}$ are positive. The expression \eqref{Xy2} is in the same form as in \cite{M1,M2}, since one can represent
\b \xi_1=\nu_1\left(y-\frac{3\kappa}{1-\nu_1^2}t -y_{10} \right),\label{xi1} \e with \b \nu_1\equiv 2\sin(l_1) \label{nu1}\e and \b y_{10}\equiv\frac{1}{\nu_1}
\ln\left(\frac{\mu_2}{\mu_1}\sqrt{\frac{(\Lambda_1-1)(\Lambda_2+1)}{(\Lambda_1+1)(\Lambda_2-1)}}\right)_{l=l_1}. \e  Note that $\gamma_1$ can be expressed from \eqref{gamma1} and $\ln \sigma $ is just a trivial additive constant. The implication of \eqref{nu1} is that the soliton parameter $\nu_1$ is restricted: $-2\leq \nu_1 \leq 2$.

However, there are more restrictions, coming from the condition that the expressions under the radical appearing in \eqref{rad1}, \eqref{gamma1} and other places should be positive. This condition is equivalent to

\begin{equation}
(\Lambda_1^2-1)(\Lambda_2^2-1)>0.
\end{equation} 

\n With \eqref{L1} - \eqref{L3} and some elementary trigonometric identities the above inequality  gives $4 \sin ^2 l_1 <1 $ which finally leads to $|\nu_1|<1.$ This is the same restriction as in \cite{M1,M2}. In terms of $l_1,$ we have $|l_1| <\pi/6$.

 In \cite{SS} it has been shown that it is possible to extend the values of the DP soliton parameters (e.g., $\nu_1>2$) in order to obtain `loop' solitons. These, however, strictly speaking, are not Schwartz class functions (not even functions) and thus such an extension is not possible in our framework.

(iii) For completeness, let us now examine the special case $l_1=0$ when in both (i) and (ii) $\Lambda_1=\Lambda_2=-\frac{1}{\sqrt{3}}$ and $\Lambda_3=\frac{2}{\sqrt{3}}$; $ A_1=A_2=-\frac{\sqrt{3}\kappa}{2}$, $A_3=\sqrt{3}\kappa$. The solution \eqref{Xy} is 

\b X(y,t)=y+ \ln\left( \frac{\frac{\sqrt{3}-1}{\sqrt{3}+1}e^{\sqrt{3}y+\frac{3\sqrt{3} }{2}\kappa t}+\frac{2+\sqrt{3}}{2-\sqrt{3}}\gamma_0}{e^{\sqrt{3}y+\frac{3\sqrt{3} }{2}\kappa t}+\gamma_0}\right) \e

\n with $\gamma_0=\frac{\mu_3}{\mu_1+\mu_2}.$ This can be matched to \eqref{Xy2}, \eqref{xi1} for $\nu_1=\sqrt{3},$ $\gamma_1=\frac{3\sqrt{3}-\sqrt{2}}{5},$ $y_{01}=\ln\left(\gamma_0\frac{\gamma_1+1}{\gamma_1-1} \right),$  $\sigma=\frac{1+\sqrt{3}}{2}.$ Therefore (iii) does not provide a new type of solution. 

\subsection{Two-soliton solution}

The two-soliton solution corresponds to two discrete eigenvalues, $\lambda_1$ and $\lambda_2$. Let us assume that they both can be chosen real, so that the Anzatz for $g$, based on the possible configuration of the poles, becomes

\b
g(y,t,\lambda)&=&\!\!1\!\!\text{I}+\frac{1}{3}\sum_{n=1}^{2}\left(\frac{\mathcal{A}_n(y,t)}{\lambda-\lambda_n}+\frac{C^{-1}\mathcal{A}_n(y,t) C}{\omega^2\lambda-\lambda_n}+\frac{C^{-2}\mathcal{A}_n(y,t)C^2}{\omega \lambda - \lambda_n} \right ) \nonumber \\ 
&\phantom{*}& \label{g2factor}  \e

\n for some residues depending on the real quantities $\mathcal{A}_1(y,t)$ and $\mathcal{A}_2(y,t)$. Again, the conditions \eqref{S1} and \eqref{S2} are automatically satisfied. The condition \eqref{S3} is equivalent to \b g(y,t,\lambda)\Gamma^{-1} g^{\dag}(y,t,- \bar{\lambda})\Gamma=\!\!1\!\!\text{I}.
\label{S3aa} \e
 
This leads to the following system of equations for $\mathcal{A}_1$, $\mathcal{A}_2$  
\b 
\left(\!\!1\!\!\text{I} - \frac{1}{3}\left(\frac{\mathcal{A}_1(y,t)}{2\lambda_1}+\frac{C^{-1}\mathcal{A}_1(y,t) C}{(\omega^2+1)\lambda_1}+\frac{C^{-2}\mathcal{A}_1(y,t)C^2}{(\omega +1) \lambda_1} \right) \right. \phantom{**********} \nonumber \\
\left. - \frac{1}{3}\left(\frac{\mathcal{A}_2(y,t)}{\lambda_1+\lambda_2}+\frac{C^{-1}\mathcal{A}_2(y,t) C}{\omega^2\lambda_1+\lambda_2}+\frac{C^{-2}\mathcal{A}_2(y,t)C^2}{\omega \lambda_1 + \lambda_2} \right)\right)\Gamma^{-1}\mathcal{A}_1^T \Gamma =0,\nonumber \e

\b
\left(\!\!1\!\!\text{I} - \frac{1}{3}\left(\frac{\mathcal{A}_1(y,t)}{\lambda_1+\lambda_2}+\frac{C^{-1}\mathcal{A}_1(y,t) C}{\omega^2\lambda_2+\lambda_1}+\frac{C^{-2}\mathcal{A}_1(y,t)C^2}{\omega\lambda_2 + \lambda_1} \right) \right. \phantom{**********} \nonumber \\
\left. - \frac{1}{3}\left(\frac{\mathcal{A}_2(y,t)}{2\lambda_2}+\frac{C^{-1}\mathcal{A}_2(y,t) C}{(\omega^2+1)\lambda_2}+\frac{C^{-2}\mathcal{A}_2(y,t)C^2}{(\omega +1) \lambda_2} \right)\right)\Gamma^{-1}\mathcal{A}_2^T \Gamma =0, \nonumber
\e which are the conditions of the vanishing of the residue of \eqref{S3aa} at $-\lambda_1$ and $-\lambda_2$. 

We seek a solution of the above matrix equations
in the form \b \mathcal{A}_1=|n\rangle \langle m|, \qquad \mathcal{A}_2=|N\rangle \langle M| \label{A12nm}\e

\n where $|n \rangle $ and $|N \rangle $  are 3-component vector-columns, and  $\langle m|$ and $\langle M | $ are 3-component vector-rows. The components of the two vectors (indexed, as usual, with indices from 1 to 3) satisfy 3 systems of two equations each. For example $n_2$ and $N_2$ are solutions of the system (for known $M_j$ and $m_j$): \begin{equation} \label{mn+MN}
\begin{split}
m_2&=\frac{m_2^2}{2\lambda_1}n_2 + \frac{\lambda_1^2 M_1 m_3 - \lambda_1 \lambda_2 M_3 m_1 + \lambda_2^2 M_2 m_2}{\lambda_1^3 + \lambda_2^3} N_2 ,\\
M_2&= \frac{\lambda_2^2 m_1 M_3 - \lambda_1 \lambda_2 m_3 M_1 + \lambda_1^2 M_2 m_2}{\lambda_1^3 + \lambda_2^3} n_2 + \frac{M_2^2}{2\lambda_2} N_2. 
\end{split}
\end{equation}

This system can be solved explicitly and the following important property can be verified: \b \frac{m_2n_2}{\lambda_1}+\frac{M_2N_2}{\lambda_2}=0.\label{prop0}\e

Similarly,  $n_3$ and $N_3$ are solutions of the system 

\begin{equation} 
\begin{split}
m_1&=\frac{m_2^2}{2\lambda_1}n_3 + \frac{-\lambda_1 \lambda_2 M_1 m_3 + \lambda_2^2 M_3 m_1 + \lambda_1^2 M_2 m_2}{\lambda_1^3 + \lambda_2^3} N_3 \\
M_1&= \frac{-\lambda_1 \lambda_2 m_1 M_3 + \lambda_1^2 m_3 M_1 + \lambda_2^2 M_2 m_2}{\lambda_1^3 + \lambda_2^3} n_3 + \frac{M_2^2}{2\lambda_2} N_3. \label{mn+MN} \end{split}
\end{equation}

With some algebra one can verify that the determinant of the above system is a complete square:

\b \Delta= \frac{\lambda_1 \lambda_2}{4(\lambda_1^3+\lambda_2^3)^2} \left( \frac{\lambda_1^3-\lambda_2^3}{\lambda_1 \lambda_2}M_2 m_2 + 2 \lambda_2 M_3 m_1 - 2\lambda_1 M_1 m_3 \right)^2,\e

\n and one can find the solutions $n_3$, $N_3$ explicitly as well:
\begin{equation}
\begin{split}
 n_3&= \frac{1}{\Delta} \left(\frac{m_1}{2\lambda_2} M_2^2 -\frac{M_1}{\lambda_1^3+\lambda_2^3}\left(\lambda_1^2 M_2 m_2 - \lambda_1 \lambda_2 M_1 m_3 + \lambda_2^2 M_3 m_1 \right) \right)                   ,                 \\
N_3&= \frac{1}{\Delta} \left(\frac{M_1}{2\lambda_1} m_2^2 -\frac{m_1}{\lambda_1^3+\lambda_2^3}\left(\lambda_1^2 M_1 m_3 - \lambda_1 \lambda_2 m_1 M_3 + \lambda_2^2 M_2 m_2 \right) \right) .\end{split} \end{equation}

 The rationale again is that when $|m \rangle $ and $|M \rangle $  are  solutions of the 'naked' spectral problem, evaluated at $\lambda=\lambda_1=\lambda(k_1)$ and $\lambda=\lambda_2=\lambda(k_2)$ correspondingly, then $|n \rangle $ and $|N \rangle $  are eigenfunctions for the two-soliton solution at these eigenvalues. And they are explicitly obtained now, since $m_j(y,t)$ are as before \eqref{m}, and $M_j$ are similar, only evaluated at $k=k_2$:

\begin{equation}
\begin{split}
M_1&=\frac{1}{\lambda_2 }\sum_{j=1}^{3}\eta_j\Lambda_j(k_2)e^{-\Lambda_j(k_2)y-A_j(k_2)t} , \\
M_2&= \sum_{j=1}^{3}\eta_j e^{-\Lambda_j(k_2)y-A_j(k_2)t} ,\\
M_3&=\lambda_2 \sum_{j=1}^{3}\frac{\eta_j}{\Lambda_j(k_2)-1}e^{-\Lambda_j(k_2)y-A_j(k_2)t} . \end{split} \label{M}
\end{equation}

\n for three new constants $\eta_j$. Next, evaluating $g$ at $\lambda=0$,
\b g(y,t,0)=\mathrm{diag} \left(1-\frac{n_1m_1}{\lambda_1}-\frac{N_1M_1}{\lambda_2},1-
\frac{n_2m_2}{\lambda_1}-\frac{N_2 M_2}{\lambda_2}, 1-\frac{n_3m_3}{\lambda_1}-\frac{N_3 M_3}{\lambda_2} \right)  , \nonumber \e

\n and noting \eqref{prop0},  we obtain

\b g(y,t,0)=\mathrm{diag} \left(1-\frac{n_1m_1}{\lambda_1}-\frac{N_1M_1}{\lambda_2},\, 1\, , 1-\frac{n_3m_3}{\lambda_1}-\frac{N_3 M_3}{\lambda_2} \right).   \e

Like in \eqref{link}, we have (the second diagonal entry is $(g)_{22}=1$)

\b  e^{X-y}\frac{\partial X}{\partial y}=+(g)_{33}= 1-\frac{n_3(y,t)m_3(y,t)}{\lambda_1}-\frac{N_3(y,t)M_3(y,t)}{\lambda_2}. \e

This differential equation can be integrated once and solved, using the fact that $m_{j}$ satisfy \eqref{meq}  and $M_j$ satisfy the same equations at $\lambda=\lambda_2$:

\begin{equation} \begin{split}
M_{1,y}=& M_1 - \lambda_2 M_3  \\
M_{2,y}=&- \lambda_2 M_1  \\
M_{3,y}= & - \lambda_2 M_2 - M_3, \end{split}
 \label{Meq}
\end{equation}

\n and $n_3$, $N_3$ are given explicitly via $m_j$, $M_j$.  The solution is

\b X(y,t)=y + \ln \left(1 + \frac{2A}{B}\right), \e

\n where \b A&=& \frac{1}{\lambda_1 \lambda_2}(\lambda_1 m_2 M_3 - \lambda_2 M_2 m_3), \\ 
B&= & \frac{\lambda_1 \lambda_2}{\lambda_1^3+ \lambda_2^3}\left( \frac{\lambda_1^3- \lambda_2^3}{\lambda_1 \lambda_2}M_2 m_2+2\lambda_2 M_3 m_1 - 2 \lambda_1 M_1 m_3 \right)  \e

Following our experience from the one-soliton solution, we take $\mu_3=0$, $\eta_3=0$ and introduce 

\b \xi_n &=& \nu_n \left(y-\frac{3\kappa}{1-\nu_n^2}t  \right) , \\
\nu_n&=&2\sin l_n \qquad \text{where} \qquad k_n=e^{il_n}, \qquad n=1,2. \e

\n The solution acquires the following form

\b X(y,t)&=&y + \ln \left(\frac{S^{(1)}_{22}+S^{(1)}_{12}e^{\xi_1}+S^{(1)}_{21}e^{\xi_2}+S^{(1)}_{11}e^{\xi_1+\xi_2}}
{S^{(2)}_{22}+S^{(2)}_{12}e^{\xi_1}+S^{(2)}_{21}e^{\xi_2}+S^{(2)}_{11}e^{\xi_1+\xi_2}}\right), \label{X2yt} \e

\begin{equation} \begin{split}
S^{(2)}_{jp}\!&=\!\left(\frac{\lambda_1^3- \lambda_2^3}{\lambda_1^3+ \lambda_2^3}+\frac{2\lambda_2^3\Lambda_j(k_1) }{(\lambda_1^3+ \lambda_2^3)(\Lambda_p(k_2)-1)} -\frac{2\lambda_1^3\Lambda_p(k_2) }{(\lambda_1^3+ \lambda_2^3)(\Lambda_j(k_1)-1)} \right) \mu_j \eta_p ,\\
S^{(1)}_{jp}\!&=\!S^{(2)}_{jp}+ 2\left( \frac{1}{\Lambda_p(k_2)-1}- \frac{1}{\Lambda_j(k_1)-1}\right) \mu_j \eta_p  . \end{split} \nonumber
 \end{equation}
With some algebra, using various identities like 
\begin{equation}
\begin{split}
&\Lambda_j^3(k_n)-\Lambda_j(k_n)-\lambda_n^3=0, \nonumber  \\
&(L_1-1)(L_2-1)(L_1^2-L_1 L_2+L_2^2-1)+2(\lambda_1^3+ \lambda_2^3) \nonumber \\
&=(L_1+1)(L_2+1)(L_2^2-L_1 L_2+L_2^2-1) ,\nonumber  \end{split}  
\end{equation}

\n for $L_1=\Lambda_j(k_1)$ and   $L_2=\Lambda_p(k_2)$,  we simplify to

\b
S^{(2)}_{jp}&=&\frac{\Lambda_j(k_1)-\Lambda_p(k_2)}{\Lambda_j(k_1)+\Lambda_p(k_2)} \, \mu_j \eta_p, \nonumber \\
S^{(1)}_{jp}&=&\frac{\Lambda_j(k_1)-\Lambda_p(k_2)}{\Lambda_j(k_1)+\Lambda_p(k_2)} \,\, \frac{\Lambda_p(k_2)+1}{\Lambda_p(k_2)-1} \, \, \frac{\Lambda_j(k_1)+1}{\Lambda_j(k_1)-1} \, \mu_j \eta_p . \nonumber
 \e

It is more convenient to introduce the following notations, like in \cite{M2}:

\b p_n&=&\sin l_n +\frac{1}{\sqrt{3}}\cos l_n = - \Lambda_1(k_n), \nonumber \\   q_n&=&\sin l_n -\frac{1}{\sqrt{3}}\cos l_n =  \Lambda_2(k_n) .\nonumber \e
Redefining \b \xi_1 \rightarrow  \xi_1&=&\nu_1 \left(y-\frac{3\kappa}{1-\nu_1^2}t  \right) - \ln \left( \frac{\mu_1}{\mu_2}\, \frac{p_1+q_2}{p_1-q_2} \,\frac{q_1+q_2}{q_1-q_2} \right) ,\\
 \xi_2 \rightarrow  \xi_2&=&\nu_2 \left(y-\frac{3\kappa}{1-\nu_2^2}t  \right) - \ln \left( \frac{\eta_1}{\eta_2}\,  \frac{q_1+p_2}{q_1-p_2} \,\frac{q_1+q_2}{q_1-q_2}\right),
 \e
\n by adding constants to the previous expressions, we finally represent the solution \eqref{X2yt} in the form 
\b X(y,t)&=&y + \ln \left(\frac{1+\rho_1e^{\xi_1}+\rho_2e^{\xi_2}+\delta \rho_1 \rho_2 e^{\xi_1+\xi_2}}{
1+e^{\xi_1}+e^{\xi_2}+\delta  e^{\xi_1+\xi_2}}\right)+ c, \label{X2yta} \e

\b
\rho_n&=& \frac{(p_n-1)(q_n-1)}{(p_n+1)(q_n+1)}=\frac{\left(1-\frac{\nu_n}{2}\right)(1-\nu_n)}{\left(1+\frac{\nu_n}{2}\right)(1+\nu_n)} \nonumber \\
\delta&=&\frac{(p_1-p_2)(q_1-q_2)(q_1-p_2)(p_1-q_2)}{(p_1+p_2)(q_1+q_2)(q_1+p_2)(p_1+q_2)}\nonumber \\
&=&\frac{(\nu_1-\nu_2)^2(\nu_1^2-\nu_1 \nu_2 +\nu_2^2 -3)}{(\nu_1+\nu_2)^2(\nu_1^2+\nu_1 \nu_2 +\nu_2^2 -3)}, \nonumber \\
c&=& \ln\frac{S^{(1)}_{22}}{S^{(2)}_{22}}= \text{const.} \nonumber \\
u(X,t)&=&X_t(y,t). \nonumber
\e

\n Now it is not difficult to spot that this form is equivalent to the 2-soliton solution from \cite{M1,M2}, up to adding appropriate constants in the definition of $\xi_1$ and $\xi_2$.  The soliton parameters $\nu_n$ are restricted as before: $| \nu_n |<1$. In \cite{SS}, for $\nu_n>2$, modified formulas are used to investigate the interactions of the `loop' 2-solitons and of a loop and a soliton. 

\section{Discussion}

In this study the dressing method has been applied  to obtain 
the one- and two-soliton solutions of the DP equation. The solutions coincide with those obtained in \cite{M1,M2} by the Hirota method. It seems that the obtained solutions are of wider class. However, the change of variables  $X(y,t)$, which should be a monotonic function in $y$ for all $t$, is enforcing heavy restrictions to the soliton parameters. It remains to give an interpretation to the case when $X(y,t)$ is not monotonic and develops singularities. It could be the case that before and after the breaking (happening at the singularity), these are still DP solutions. Then it should be established if such solutions are conservative or not.

 The peakon solutions (peaked solitons) appear in the limit $\kappa
\to 0$ \cite{L-S}. The limit when $\nu_1 \to 1$, for the 1-soliton case (to the one-peakon) is explained in \cite{M1}.

The dressing factor for the $\mathcal{N}$- soliton solution is \b
g(y,t,\lambda)&=&\!\!1\!\!\text{I}+\frac{1}{3}\sum_{n=1}^{\mathcal{N}}\left(\frac{\mathcal{A}_n(y,t)}{\lambda-\lambda_n}+\frac{C^{-1}\mathcal{A}_n(y,t) C}{\omega^2\lambda-\lambda_n}+\frac{C^{-2}\mathcal{A}_n(y,t)C^2}{\omega \lambda - \lambda_n} \right ) \nonumber \\
\label{Ns} \e and all steps can be followed. However, the technical level increases significantly.

 An important question is whether the construction \eqref{gfactor} and, more generally, \eqref{Ns} with a rank one residues $\mathcal{A}_n$ covers all possible $\mathcal{N}$-soliton solutions of the DP equation. So far there are no indications in the existing literature for other types of solitons. Nevertheless there are several other possibilities for the choice of the dressing factor. This can potentially lead to new types of solitons and remains to be studied in the future. Such choices include (but are not limited to)

(i) higher rank residue $\mathcal{A}_1$,

(ii) dressing factor with 6 poles, i.e. with additional poles at $-\lambda_1$, $-\omega \lambda_1$ and $-\omega^2 \lambda_1$,

(iii) dressing factor with two eigenvalues which are $\lambda_1$ and the complex conjugate $\lambda_2=\bar{\lambda}_1$, then $\mathcal{A}_2=\bar{\mathcal{A}}_1$ etc. Such solitons exist for the Tzitzeica equation \cite{BCG1,BCG2}. However, they include oscillating $\sin $-$\cos$ factors and then $X(y,t)$ is unlikely to be monotonic.

A two-component integrable version of the DP equation is constructed in \cite{GH}. Its spectral properties have some similarity to those of the DP equation and it should be possible to apply the methods from \cite{CIL} and the present paper to this equation. Note that the multidimensional versions of the DP
equation are in general non-integrable, but admit singular
(peakon-type) solutions \cite{HS}.

\section*{Acknowledgments}

The support of the FWF Project I544-N13 ``Lagrangian kinematics of water waves'' of the Austrian Science Fund is gratefully acknowledged. RI was also
supported by a {\it Seed funding} grant from Dublin Institute of Technology for a project in association with the Environmental Sustainability and Health Institute,  Dublin.

The authors are grateful to Prof. V.~S.~Gerdjikov for many valuable discussions and to two ananymous referees for some very important suggestions.


\end{document}